\documentclass[11pt,preprintnumbers,titlepage,nofootinbib]{revtex4-2}%
\usepackage[latin9]{inputenc}
\usepackage{amsmath}
\usepackage{amssymb}
\usepackage{graphicx}
\usepackage{wasysym}
\usepackage[unicode=true,  bookmarks=false,  breaklinks=false,pdfborder={0 0
1},backref=section,colorlinks=false]{hyperref}
\usepackage{array}
\usepackage{multirow}
\usepackage{wasysym}
\usepackage{wasysym}
\usepackage{amsfonts}
\usepackage{subfigure}
\usepackage{cleveref}
\usepackage{listings}
\usepackage{xcolor}
\usepackage{array}
\usepackage{url}
\usepackage{color}%
\hypersetup{
colorlinks,linkcolor=blue,citecolor=blue,urlcolor=blue}
\makeatletter

\@ifundefined{textcolor}{}{
\definecolor{BLACK}{gray}{0}
\definecolor{WHITE}{gray}{1}
\definecolor{RED}{rgb}{1,0,0}
\definecolor{GREEN}{rgb}{0,1,0}
\definecolor{BLUE}{rgb}{0,0,1}
\definecolor{CYAN}{cmyk}{1,0,0,0}
\definecolor{MAGENTA}{cmyk}{0,1,0,0}
\definecolor{YELLOW}{cmyk}{0,0,1,0}
}

\makeatother
\begin{document}
\preprint{CTP-SCU/2025015}
\title{Schwarzschild Black Holes Immersed in Born-Infeld Magnetic Fields and Their
Observational Signatures}
\author{Yiqian Chen$^{b}$}
\email{chenyiqian@ucas.ac.cn}
\author{Guangzhou Guo$^{c}$}
\email{guogz@sustech.edu.cn}
\author{Benrong Mu$^{a}$}
\email{benrongmu@cdutcm.edu.cn}
\author{Peng Wang$^{d}$}
\email{pengw@scu.edu.cn}
\affiliation{$^{a}$Center for Joint Quantum Studies, College of Medical Technology, Chengdu
University of Traditional Chinese Medicine, Chengdu, 611137, PR China }
\affiliation{$^{b}$School of Fundamental Physics and Mathematical Sciences, Hangzhou
Institute for Advanced Study, University of Chinese Academy of Sciences,
Hangzhou, 310024, China}
\affiliation{$^{c}$Department of Physics, Southern University of Science and Technology,
Shenzhen, 518055, China}
\affiliation{$^{d}$College of Physics, Sichuan University, Chengdu, 610064, China}

\begin{abstract}
We investigate the influence of Born-Infeld (BI) nonlinear electrodynamics on
magnetic field configurations, photon orbits, and black hole shadows for
Schwarzschild black holes immersed in magnetic fields. Assuming that the BI
magnetic fields are asymptotically uniform and aligned with the polar axis, we
solve the nonlinear magnetic field equations numerically using pseudospectral
methods. Our analysis shows that nonlinear electromagnetic effects become
prominent near the event horizon, particularly in the polar regions, where the
magnetic field strength is significantly enhanced. This enhancement leads to
closed photon orbits on the meridional plane becoming prolate, with noticeable
stretching along the polar axis. Simulations of black hole images reveal that,
at high observer inclinations, the shadow, which is circular in the Maxwell
limit, becomes increasingly elongated along the polar direction as the
nonlinear effects increase.

\end{abstract}
\maketitle
\tableofcontents

{}

\section{Introduction}

\label{sec:Introduction}

The recent release of the first horizon-scale image of the supermassive black
hole M87* by the Event Horizon Telescope (EHT) collaboration represents a
major milestone in black hole astrophysics
\cite{Akiyama:2019bqs,Akiyama:2019brx,Akiyama:2019cqa,Akiyama:2019eap,Akiyama:2019fyp,Akiyama:2019sww}%
. This achievement was followed by the publication of the image of Sgr A*, the
supermassive black hole at the center of the Milky Way
\cite{EventHorizonTelescope:2022xnr,EventHorizonTelescope:2022vjs,EventHorizonTelescope:2022wok,EventHorizonTelescope:2022exc,EventHorizonTelescope:2022urf,EventHorizonTelescope:2022xqj}%
, marking the beginning of a new era in testing general relativity in the
strong-field regime. Black hole images are typically characterized by a
central dark region---referred to as the \textquotedblleft
shadow\textquotedblright---surrounded by a bright emission ring associated
with closed photon orbits. These features are interpreted as direct
consequences of strong gravitational lensing, which significantly deflects
light near black holes
\cite{Synge:1966okc,Bardeen:1972fi,Bardeen:1973tla,Bozza:2009yw}.

The finite resolution of black hole images leaves room for alternative models
that can reproduce the observed features within current observational
uncertainties. Utilizing EHT data, it is possible to place constraints on the
parameters of models that deviate from the Kerr geometry, thereby enhancing
our understanding of the black hole environment
\cite{Li:2013jra,Tsukamoto:2014tja,Kumar:2018ple,Bambi:2019tjh,Afrin:2022ztr,Wang:2025fmz}. In recent years, the black
hole shadow and related optical phenomena have been extensively studied across
a variety of theoretical frameworks, including hairy black holes
\cite{Cunha:2015yba,Cunha:2016bpi,Khodadi:2020jij,Gan:2021pwu,Gan:2021xdl,Ghosh:2023kge,Fernandes:2024ztk,Wang:2024lte}%
, string inspired black holes
\cite{Amarilla:2011fx,Guo:2019lur,Zhu:2019ura,Kumar:2020hgm}, fuzzball
\cite{Bacchini:2021fig}, exotic ultra-compact objects
\cite{Cunha:2017wao,Cunha:2018acu,Shaikh:2018oul,Shaikh:2019itn,Shaikh:2019jfr,Wielgus:2020uqz,Peng:2021osd,Huang:2024bbs}%
, naked singularities
\cite{Joshi:2020tlq,Dey:2020bgo,Chen:2023trn,Chen:2023knf,Deliyski:2024wmt},
and modified gravity theories
\cite{Amarilla:2010zq,Ayzenberg:2018jip,Wang:2018prk,Ma:2019ybz,Guo:2020zmf,Wei:2020ght,Zeng:2020dco,Addazi:2021pty,Wang:2022yvi,Liu:2024lve}%
.

To further refine the image of M87*, the EHT collaboration employed GRMHD
simulations to reconstruct its polarized structure \cite{Akiyama:2021qum}. The
results from these simulations support the existence of a magnetically
arrested disk surrounding the black hole
\cite{Igumenshchev:2003rt,Narayan:2003by,Akiyama:2021tfw}. Moreover, strong
magnetic fields have been detected near Sgr A*, the supermassive black hole at
the center of the Milky Way \cite{Eatough:2013nva}. Motivated by such
observations, the influence of magnetic fields on black hole images has become
an active area of study \cite{Junior:2021svb,Wang:2021ara,Hou:2022eev}. In the
extreme environments near black holes---where magnetic fields can reach
immense strengths---the assumption that the electromagnetic sector is
accurately described by Maxwell's linear electrodynamics may no longer hold.
This motivates the investigation of alternative nonlinear electrodynamics
theories. Indeed, the impact of nonlinear electrodynamics on black hole
shadows has been explored in various works
\cite{Atamurotov:2015xfa,Stuchlik:2019uvf,Allahyari:2019jqz,Zare:2024dtf,AraujoFilho:2024xhm}. In particular, black
holes immersed in external magnetic fields provide a natural setting to probe
nonlinear electromagnetic effects, especially in relation to observational
signatures detectable by current and future astronomical facilities
\cite{Hu:2020usx,Zhong:2021mty,He:2022opa}.

Among various nonlinear electrodynamics theories, Born-Infeld (BI)
electrodynamics---originally inspired by string theory---stands out for its
role in describing the low-energy dynamics of D-branes
\cite{Fradkin:1985qd,Tseytlin:1986ti,Salazar:1987ap,Wiltshire:1988uq}. One of
the key features of BI electrodynamics is the introduction of a maximal
electric field, which regularizes the electrostatic self-energy of point
charges \cite{Born:1934gh}. Recent integrations of BI electrodynamics into
gravity frameworks have generated significant theoretical interest
\cite{Cai:2004eh,Dey:2004yt,Fernando:2006gh,Miskovic:2008ck,Gunasekaran:2012dq,Zou:2013owa,Tao:2017fsy,Wang:2018xdz,Gan:2019jac,Wang:2019kxp,Wang:2020ohb}%
. Specifically, for a Schwarzschild black hole immersed in an asymptotically
uniform magnetic field, perturbative solutions for the BI magnetic field and
the resulting black hole shadow have been obtained by considering only the
lowest-order corrections due to BI nonlinearities
\cite{Bokulic:2019kcc,He:2022opa}. However, the full non-perturbative solution
for the BI magnetic field in the Schwarzschild background remains unknown.
This work addresses this gap by numerically solving the BI field equations
with boundary conditions ensuring asymptotic uniformity, employing
pseudospectral methods for accuracy and efficiency.

The remainder of this paper is organized as follows. In Sec. \ref{sec:setup},
we numerically compute the BI magnetic field configuration in the
Schwarzschild black hole background and analyze its behavior with varying
nonlinearity parameter. In Sec. \ref{sec:Optical-appearance}, we investigate
the resulting closed photon orbits and simulate black hole images for
different values of the nonlinearity parameter and observer inclination.
Finally, we summarize our main findings in Sec. \ref{sec:Conclusion}. We set
$G=c=4\pi\epsilon_{0}=1$ throughout the paper.

\section{Born-Infeld Magnetic Field}

\label{sec:setup}

In this section, we consider an asymptotically uniform BI magnetic field in
the Schwarzschild black hole background, aligned along the $z$-axis. The BI
magnetic field is governed by the Lagrangian density \cite{Born:1934gh},
\begin{equation}
\mathcal{L}\left(  F\right)  =-\frac{4}{a}\left(  1-\sqrt{1+\frac{aF}{2}%
}\right)  , \label{eq:BILag}%
\end{equation}
where $a$ is the nonlinearity parameter, $F=F^{\mu\nu}F_{\mu\nu}$, and
$F_{\mu\nu}=\partial_{\mu}A_{\nu}-\partial_{\nu}A_{\mu}$ is the
electromagnetic field strength tensor. In the limit $a\rightarrow0$, the BI
theory reduces to Maxwell electrodynamics. Neglecting the backreaction of the
BI field, the background spacetime is described by the Schwarzschild metric,
\begin{equation}
ds^{2}=-f\left(  r\right)  dt^{2}+\frac{1}{f\left(  r\right)  }dr^{2}%
+r^{2}\left(  d\theta^{2}+\text{sin}^{2}\theta d\varphi^{2}\right)  ,
\end{equation}
where the metric function is $f\left(  r\right)  =1-2M/r$, with $M$ denoting
the mass of the black hole.

By varying the Lagrangian in Eq. $\left(  \ref{eq:BILag}\right)  $ with
respect to the electromagnetic field $A_{\mu}$, one obtains the equation of
motion,
\begin{equation}
\partial_{\mu}\left(  \frac{\sqrt{-g}F^{\mu\nu}}{\sqrt{1+aF/2}}\right)  =0.
\label{eq:EOM}%
\end{equation}
For an asymptotically uniform magnetic field, we adopt the ansatz,
\begin{equation}
A=u\left(  r,\theta\right)  d\varphi, \label{eq:ansatz}%
\end{equation}
where $A_{t}=0$, as we consider a purely magnetic configuration without
electric charge. In the Maxwell limit $a=0$, the equation of motion $\left(
\ref{eq:EOM}\right)  $ admits the analytical solution \cite{Wald:1974np},
\begin{equation}
u\left(  r,\theta\right)  =\frac{1}{2}B_{0}r^{2}\sin^{2}\theta.
\label{eq:linerMF}%
\end{equation}

To numerically solve the partial differential equation in Eq. $\left(
\ref{eq:EOM}\right)  $, appropriate boundary conditions for the ansatz
$u\left(  r,\theta\right)  $ must be specified. Since the BI magnetic field is
assumed to become uniform at spatial infinity, the solution should be
symmetric with respect to reflection across the equatorial plane and approach
the asymptotic linear solution given in Eq. $\left(  \ref{eq:linerMF}\right)
$ in the far-field regime \cite{Bokulic:2019kcc}. Accordingly, the function
$u\left(  r,\theta\right)  $ should satisfy the following boundary
conditions,
\begin{equation}
\partial_{\theta}u\left(  r,\theta\right)  |_{\theta=\pi/2}=0,\qquad
\lim_{r\rightarrow\infty}u\left(  r,\theta\right)  /r^{2}=B_{0}\sin^{2}\left(
\theta\right)  /2. \label{eq:BC1}%
\end{equation}
Additionally, a power series expansion of $u\left(  r,\theta\right)  $ near
$\theta=0$ and $r=0$ yields the conditions,
\begin{equation}
u\left(  r,\theta\right)  |_{\theta=0}=0,\qquad\partial_{r}u\left(
r,\theta\right)  |_{r=0}=0. \label{eq:BC2}%
\end{equation}

In this work, we employ pseudospectral methods to numerically solve the
nonlinear partial differential equation given in Eq. $\left(  \ref{eq:EOM}%
\right)  $, subject to the boundary conditions specified in Eqs. $\left(
\ref{eq:BC1}\right)  $ and $\left(  \ref{eq:BC2}\right)  $. Pseudospectral
methods are a well-established approach for solving partial differential
equations \cite{boyd2001chebyshev}. The function $u\left(  r,\theta\right)  $
is approximated by a finite sum of basis functions as%
\begin{equation}
u\left(  r,\theta\right)  =%
{\displaystyle\sum\limits_{i=0}^{N_{x}-1}}
{\displaystyle\sum\limits_{j=0}^{N_{\theta}-1}}
\alpha_{ij}T_{i}\left(  \tilde{r}\right)  \cos\left(  2j\theta\right)  ,
\label{eq:sexpansion}%
\end{equation}
where $N_{x}$ and $N_{\theta}$ are the spectral resolutions in the radial and
angular directions, respectively, $T_{i}\left(  \tilde{r}\right)  $ denotes
the Chebyshev polynomials, and $\alpha_{ij}$ are the spectral coefficients.

To compactify the radial domain, we introduce a coordinate transformation,
\begin{equation}
\tilde{r}=\frac{\sqrt{r^{2}-r_{H}^{2}}-r_{H}}{\sqrt{r^{2}-r_{H}^{2}}+r_{H}},
\end{equation}
which maps the event horizon $r_{H}=2M$ to $\tilde{r}=-1$ and spatial infinity
to $\tilde{r}=1$. We choose a spectral resolution of $\left(  N_{x},N_{\theta
}\right)  =\left(  22,22\right)  $ to balance accuracy and computational
efficiency. The spectral expansion in Eq. $\left(  \ref{eq:sexpansion}\right)
$ is substituted into the equation of motion, and the resulting expressions
are evaluated at Gauss-Chebyshev collocation points. This procedure transforms
the partial differential equation into a system of algebraic equations for the
coefficients $\alpha_{ij}$, which we solve using a standard Newton-Raphson
iteration. At each iteration, the resulting linear system is solved using the
built-in LinearSolve function in Mathematica.

Once $u\left(  r,\theta\right)  $ (and consequently $F_{\mu\nu}$) is
determined, the magnetic field $B$ measured by an observer with four-velocity
$U^{\mu}$ is given by\textbf{ }%
\begin{equation}
B^{i}\equiv-\frac{1}{2}\epsilon^{i\nu\rho\sigma}U_{\nu}F_{\rho\sigma},
\label{eq: def}%
\end{equation}
where $\epsilon^{\mu\nu\rho\sigma}\equiv-\tilde{\epsilon}^{\mu\nu\rho\sigma
}/\sqrt{-g}$ is the Levi-Civita tensor, $\tilde{\epsilon}^{\mu\nu\rho\sigma}$
denotes the permutation symbol, and the index $i$ denotes spatial components.
Specifically, we consider a static observer defined by the following
orthonormal tetrad:%
\begin{equation}
e^{\left(  t\right)  }=\sqrt{1-\frac{2M}{r}}dt,\text{ }e^{\left(  r\right)
}=\frac{1}{\sqrt{1-\frac{2M}{r}}}dr,\text{ }e^{\left(  \theta\right)
}=rd\theta,\text{ }e^{\left(  \varphi\right)  }=r\sin\theta d\varphi\text{.}%
\end{equation}
In this tetrad frame, the components of the magnetic field are expressed as
\begin{equation}
B^{\left(  r\right)  }=\frac{1}{r^{2}\sin\theta}\partial_{\theta}u\left(
r,\theta\right)  ,\;B^{\left(  \theta\right)  }=-\frac{\sqrt{1-2M/r}}%
{r\sin\theta}\partial_{r}u\left(  r,\theta\right)  ,\;B^{\left(
\varphi\right)  }=0, \label{eq: Bcomponents}%
\end{equation}
indicating that $B^{\left(  \theta\right)  }=0$ at the event horizon. This
implies that the magnetic field lines are perpendicular to the horizon
surface. As a consistency check, we compare our numerical results for small
$a$ with the perturbative solutions for the BI magnetic field presented in
\cite{Bokulic:2019kcc,He:2022opa}, and find good agreement.

\begin{figure}[ptb]
\includegraphics[width=1\textwidth]{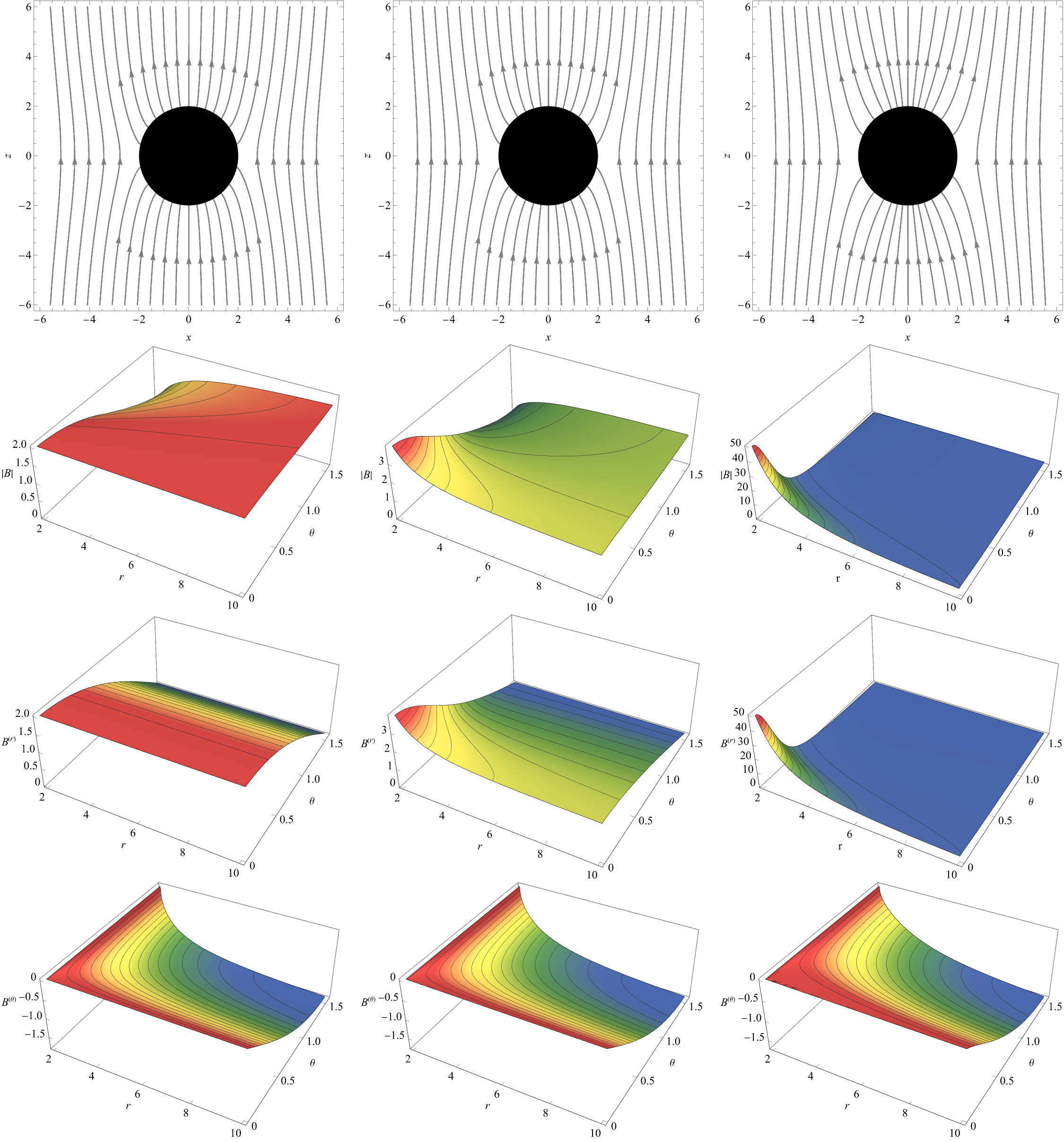}\caption{Magnetic field
configurations around a Schwarzschild black hole for different values of the
nonlinearity parameter: $a=0$ (\textbf{Left}), $a=0.5$ (\textbf{Middle}), and
$a=2$ (\textbf{Right}), with $M=1$ and $B_{0}=2$. \textbf{Top Row}: Magnetic
field lines in the $\varphi=0$ plane (spanned by $x$ and $z$). The overall
structure is similar across different $a$, but line density increases near the
polar axis close to the horizon as $a$ increases. \textbf{Second Row}:
Magnitude of the magnetic field $\left\vert B\right\vert $, which shows
enhancement near the polar region for nonzero $a$. \textbf{Third Row}: Radial
tetrad component $B^{\left(  r\right)  }$. Nonlinear effects primarily
influence $B^{\left(  r\right)  }$. \textbf{Bottom Row}: Polar tetrad
component $B^{\left(  \theta\right)  }$, which remains similar across $a$.}%
\label{Fig: MagFieldConfiguration}%
\end{figure}

Fig. \ref{Fig: MagFieldConfiguration} illustrates magnetic field
configurations around a Schwarzschild black hole for three representative
values of the nonlinearity parameter: $a=0$ (left column), $a=0.5$ (middle
column), and $a=2$ (right column). For simplicity, we set $M=1$ and $B_{0}=2$.
The top row shows magnetic field lines in the $\varphi=0$ plane, spanned by
the Cartesian coordinates $x$ and $z$. While the overall structure remains
similar across different $a$ values, the field line density near the polar
axis close to the event horizon increases noticeably with $a$. The second row
presents the magnitude of the magnetic field $\left\vert B\right\vert $, which
generally decreases with increasing polar angle $\theta$ near the black hole.
In the Maxwell limit, $\left\vert B\right\vert $ remains nearly constant near
the polar axis, whereas in the BI cases, it exhibits a pronounced enhancement
toward the horizon. This enhancement intensifies with increasing $a$,
consistent with the increased field line density. The third and bottom rows
show the radial tetrad component $B^{\left(  r\right)  }$ and polar tetrad
component $B^{\left(  \theta\right)  }$, respectively. As expected,
$B^{\left(  \theta\right)  }$ vanishes precisely on the event horizon for all
values of $a$, in agreement with the field lines shown in the top row and the
analytic result in Eq. $\left(  \ref{eq: Bcomponents}\right)  $. Near the
equatorial plane, $B^{\left(  \theta\right)  }$ dominates the field magnitude,
while $B^{\left(  r\right)  }$ becomes the leading component away from the
plane. Moreover, the distribution of $B^{\left(  \theta\right)  }$ remains
largely unchanged across different $a$, indicating that nonlinear
electromagnetic effects predominantly influence the radial component
$B^{\left(  r\right)  }$.

\section{Observational Signatures}

\label{sec:Optical-appearance}

In this section, we first analyze null geodesics in the effective geometry
induced by BI magnetic fields. We then investigate optical appearances of
Schwarzschild black holes immersed in BI magnetic fields, illuminated by
either a luminous celestial sphere or a geometrically and optically thin
accretion disk. Throughout this section, we set $M=1$ and $B_{0}=2$.

\subsection{Light Ray}

\label{subsec:Light-ray}

Nonlinear electrodynamics theories introduce self-interactions of the
electromagnetic field, which in turn modify the propagation of photons.
Consequently, photons no longer follow null geodesics with respect to the
background metric $g_{\mu\nu}$. To describe the trajectories of light rays in
the presence of a nonlinear magnetic field, an effective geometry must be
employed. As shown in \cite{Novello:1999pg,He:2022opa}, the effective metric
$G_{\mu\nu}$ for BI electrodynamics is given by
\begin{equation}
G^{\mu\nu}=\partial_{F}\mathcal{L}\left(  F\right)  g^{\mu\nu}-4\partial
_{F}^{2}\mathcal{L}\left(  F\right)  F_{\;\alpha}^{\mu}F^{\alpha\nu},
\label{eq:EffMetric}%
\end{equation}
with inverse $G^{\lambda\mu}$ satisfying $G^{\lambda\mu}G_{\mu\nu}=\delta
_{\nu}^{\lambda}.$

Accordingly, the motion of a photon is governed by the geodesic equation,
\begin{equation}
\frac{d^{2}x^{\mu}}{d\lambda^{2}}+\Gamma_{\rho\sigma}^{\mu}\frac{dx^{\rho}%
}{d\lambda}\frac{dx^{\sigma}}{d\lambda}=0, \label{eq:geoEq}%
\end{equation}
where $\lambda$ is an affine parameter, and the Christoffel symbols
$\Gamma_{\rho\sigma}^{\mu}$ are computed with respect to the effective metric
$G_{\mu\nu}$. The photon trajectory also satisfies the Hamiltonian constraint
$\mathcal{H}=G_{\mu\nu}p^{\mu}p^{\nu}/2=0$, where $p^{\mu}=dx^{\mu}/d\lambda$.
Given that the spacetime admits two Killing vectors, $K^{\mu}=\left(
1,0,0,0\right)  $ and $R^{\mu}=\left(  0,0,0,1\right)  $, the following
conserved quantities arise,
\begin{align}
E  &  =-K_{\mu}p^{\mu}=-q_{t},\nonumber\\
L  &  =R_{\mu}p^{\mu}=q_{\varphi},
\end{align}
where the dual momentum vector $q_{\mu}\equiv G_{\mu\nu}p^{\nu}$. The
quantities $E$ and $L$ correspond to the photon's conserved energy and angular
momentum, respectively.

When light rays are confined to the equatorial plane with $\theta=\pi/2$, the
radial component of the geodesic equation takes the form
\begin{equation}
-\frac{G_{tt}\left(  r,\pi/2\right)  G_{rr}\left(  r,\pi/2\right)  }{L^{2}%
}\left(  \frac{dr}{d\lambda}\right)  ^{2}=\frac{E^{2}}{L^{2}}-V_{\text{eff}%
}\left(  r\right)  , \label{eq:radial=000020EOM}%
\end{equation}
where the effective potential is defined as $V_{\text{eff}}\left(  r\right)
=-G_{tt}\left(  r,\pi/2\right)  /G_{\varphi\varphi}\left(  r,\pi/2\right)  $.
Unstable circular photon orbits on the equatorial plane at radius $r=r_{c}$
are determined by the conditions
\begin{equation}
V_{\text{eff}}\left(  r_{c}\right)  =E^{2}/L^{2},\text{ }V_{\text{eff}%
}^{\prime}\left(  r_{c}\right)  =0,\text{ }V_{\text{eff}}^{\prime\prime
}\left(  r_{c}\right)  <0,
\end{equation}
where primes denote derivatives with respect to the radial coordinate $r$. Due
to axisymmetry, light rays may also propagate within the meridional plane
defined by $\varphi=0$. On this plane, the magnetic field lines are symmetric
about the $z$-axis. As a result, closed photon orbits in the meridional plane
are expected to exhibit reflection symmetry about the $z$-axis, implying
$p^{r}=0$ at $\theta=0$. To identify such closed orbits, we numerically
integrate Eq. $\left(  \ref{eq:geoEq}\right)  $, initiating the light ray at
$\left(  r,\theta\right)  =\left(  r_{c},0\right)  $, and search for the
critical value of $r_{c}$ that results in a closed trajectory.

\begin{figure}[tb]
\includegraphics[width=0.8\textwidth]{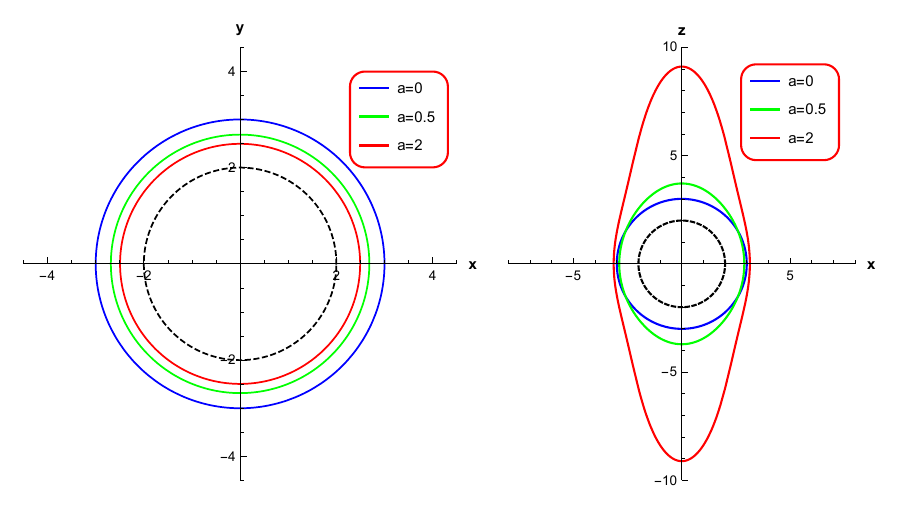}\caption{Closed photon
orbits on the equatorial plane with $\theta=\pi/2$ (\textbf{Left}) and the
meridional plane with $\varphi=0$ (\textbf{Right}) for various values of the
parameter $a$. The dashed black circle indicates the event horizon. For $a=0$
(Maxwell case), closed orbits are circular. With increasing $a$, closed orbits
shrink on the equatorial plane. On the meridional plane, closed orbits become
prolate, stretching along the $z$-axis. The closed orbit for $a=0.5$ is
slightly compressed along the $x$-axis relative to $a=0$, while for $a=2$, it
is slightly stretched.}%
\label{fig:closed=000020orbits}%
\end{figure}

Fig. \ref{fig:closed=000020orbits} displays closed orbits of photons for
different values of the BI nonlinearity parameter $a$. The left panel shows
these orbits on the equatorial plane, while the right panel illustrates them
on the meridional plane. In both panels, the dashed black circle represents
the event horizon. For $a=0$, which corresponds to the linear Maxwell theory,
the closed orbits are circular in both planes. As the parameter $a$ increases,
indicating a stronger nonlinear electromagnetic effect, the closed orbits
deform. On the equatorial plane, the orbits shrink inwards, becoming smaller
circles. On the meridional plane, the deviation from circularity becomes more
pronounced, stretching along the $z$-axis while remaining symmetric with
respect to the $z$-axis. In particular, the orbit corresponding to $a=2$
develops a noticeable elongation along the $z$-axis, highlighting the
significant impact of nonlinear electromagnetic effects on photon trajectories
in strong-field regions.

To image black holes, we employ a numerical backward ray-tracing method to
trace light rays from observers to emitting sources. Initial conditions for
Eq. $\left(  \ref{eq:geoEq}\right)  $ are set by considering the photon
$4$-momentum measured by a static observer located at $x_{o}^{\mu}=\left(
t_{o},r_{o},\theta_{o},\varphi_{o}\right)  ^{\ref{ft:1}}$ \footnotetext[1]%
{\label{ft:1} Note that a different observer, one static with respect to the
effective geometry, was considered in \cite{He:2022opa}. This choice leads to
shadows stretched along the equatorial plane, rather than the polar
direction.},
\begin{equation}
p^{\left(  t\right)  }=p^{t}\sqrt{-g_{tt}\left(  x_{o}\right)  },\quad
p^{\left(  r\right)  }=p^{r}\sqrt{g_{rr}\left(  x_{o}\right)  },\quad
p^{\left(  \theta\right)  }=p^{\theta}\sqrt{g_{\theta\theta}\left(
x_{o}\right)  },\quad p^{\left(  \varphi\right)  }=p^{\varphi}\sqrt
{g_{\varphi\varphi}\left(  x_{o}\right)  }.
\end{equation}
The observation angles $\left(  \alpha,\beta\right)  $ are defined via the
photon's local momentum components as in \cite{Cunha:2018acu},
\begin{equation}
\sin\alpha=\frac{p^{\left(  \theta\right)  }}{|\overrightarrow{P}|},\quad
\tan\beta=\frac{p^{\left(  \varphi\right)  }}{p^{\left(  r\right)  }},
\end{equation}
where $|\overrightarrow{P}|=\sqrt{p^{\left(  r\right)  2}+p^{\left(
\theta\right)  2}+p^{\left(  \varphi\right)  2}}$. The photon momentum
$p^{\mu}\left(  x_{o}\right)  $ serves as the initial data for tracing
geodesics backward to the emission region. By imposing the Hamiltonian
constrain $\mathcal{H}=0$ and normalizing the local spatial momentum to
$|\overrightarrow{P}|=1$, the initial momentum is uniquely determined by the
observation angles $\alpha$ and $\beta$. The observer's image plane
coordinates $\left(  x,y\right)  $ are defined as
\begin{equation}
x\equiv-r_{o}\beta,\quad y\equiv r_{o}\alpha,
\end{equation}
where the zero observation angles $\left(  0,0\right)  $ correspond to the
direction pointing toward the black hole center.

\subsection{Celestial Sphere}

\label{subsec:Celestial-sphere}

We now investigate images of a Schwarzschild black hole immersed in BI
magnetic fields and illuminated by a luminous celestial sphere. The celestial
sphere is assumed to have a radius of $r_{\text{CS}}=100$ and is concentric
with the black hole. A static observer is located at $r_{o}=50$, with a field
of view spanning $\pi/3$. To construct the images, we numerically integrate
$2000\times2000$ null geodesics from the observer's location, continuing each
ray until it either intersects the celestial sphere or falls into the event
horizon. The celestial sphere is divided into four quadrants, each assigned a
distinct color (see Fig. 2 in \cite{Guo:2022muy}). Additionally, we overlay a
grid of black lines representing constant longitude and latitude, with
adjacent lines separated by $\pi/18$.

\begin{figure}[ptb]
\includegraphics[width=0.3\textwidth]{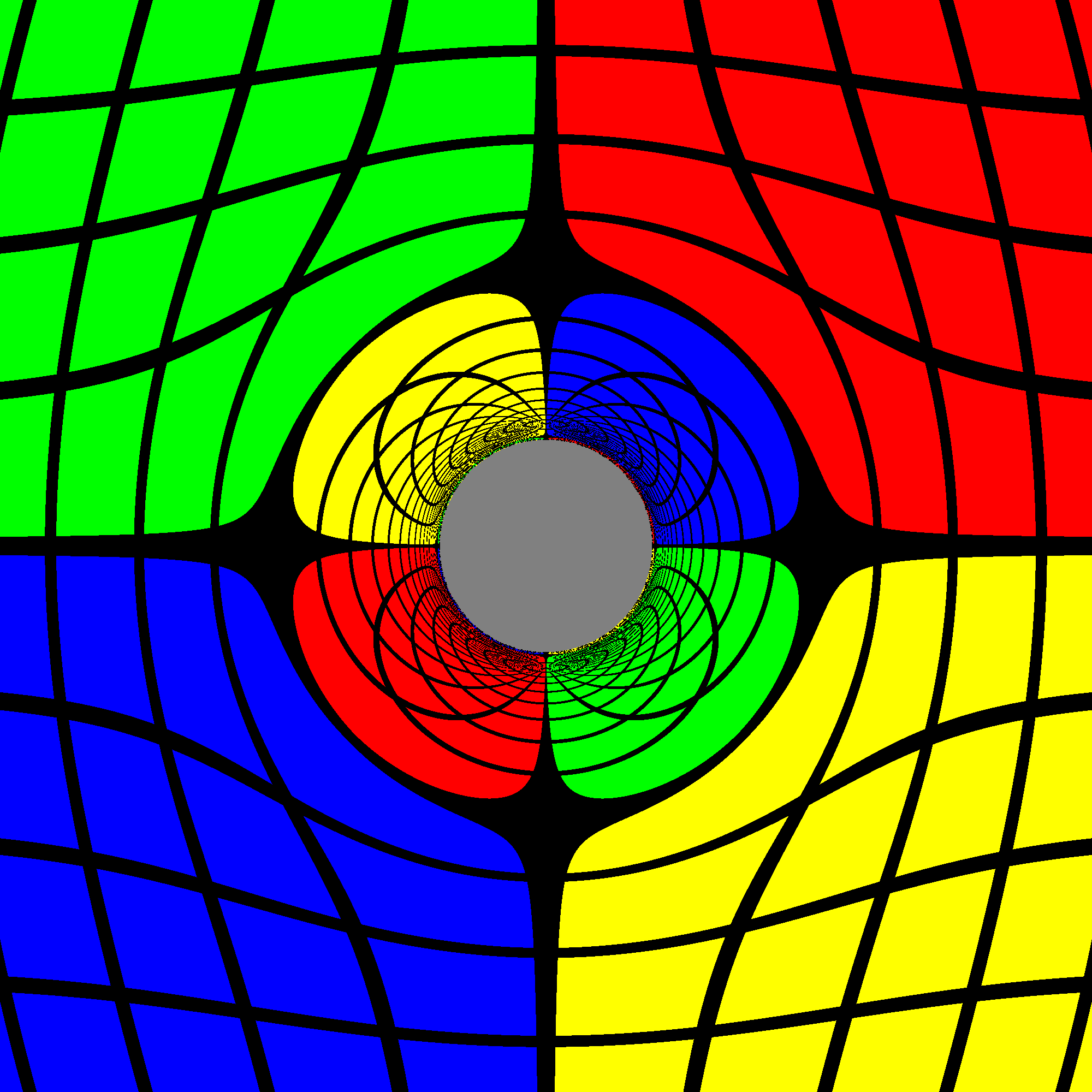}
\includegraphics[width=0.3\textwidth]{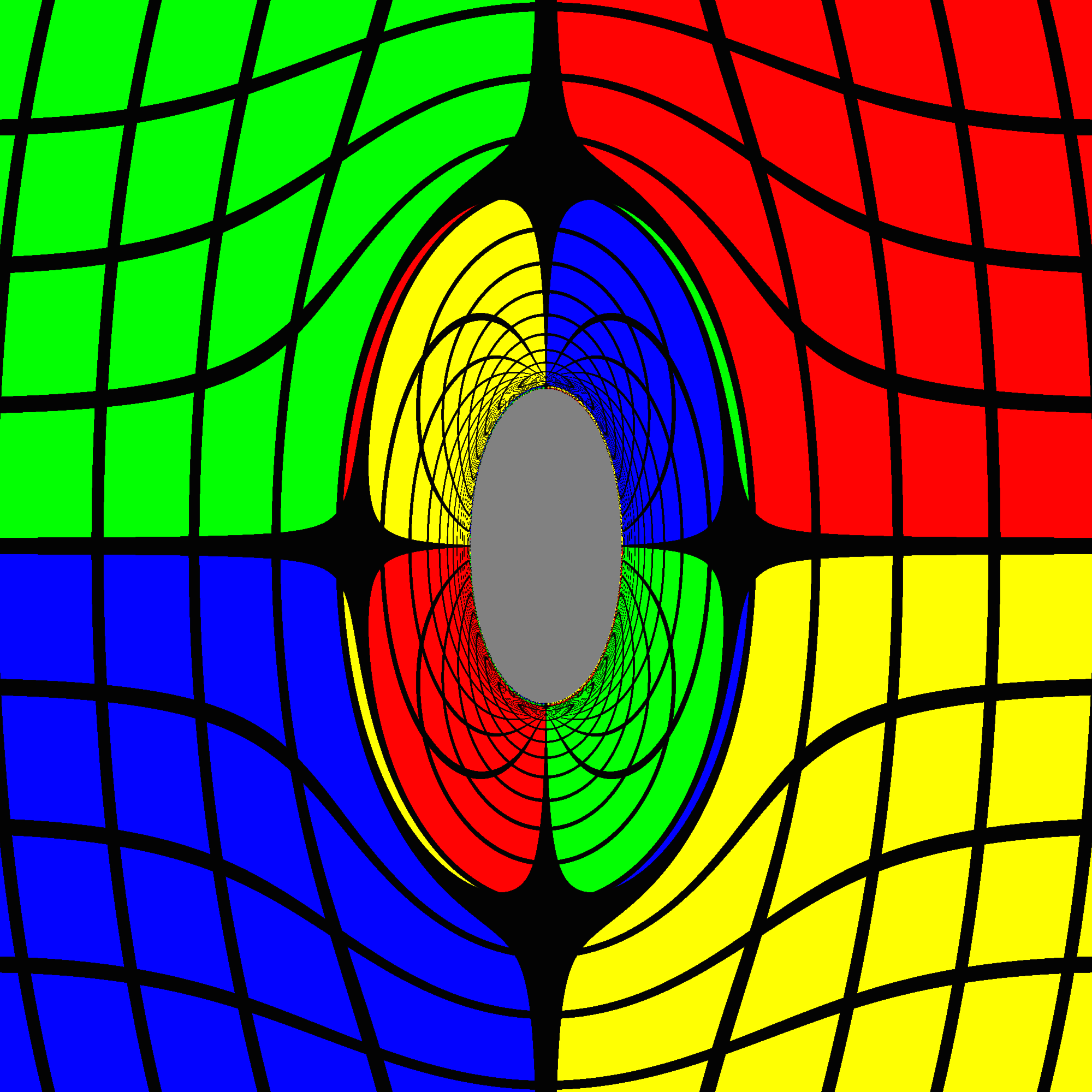}
\includegraphics[width=0.3\textwidth]{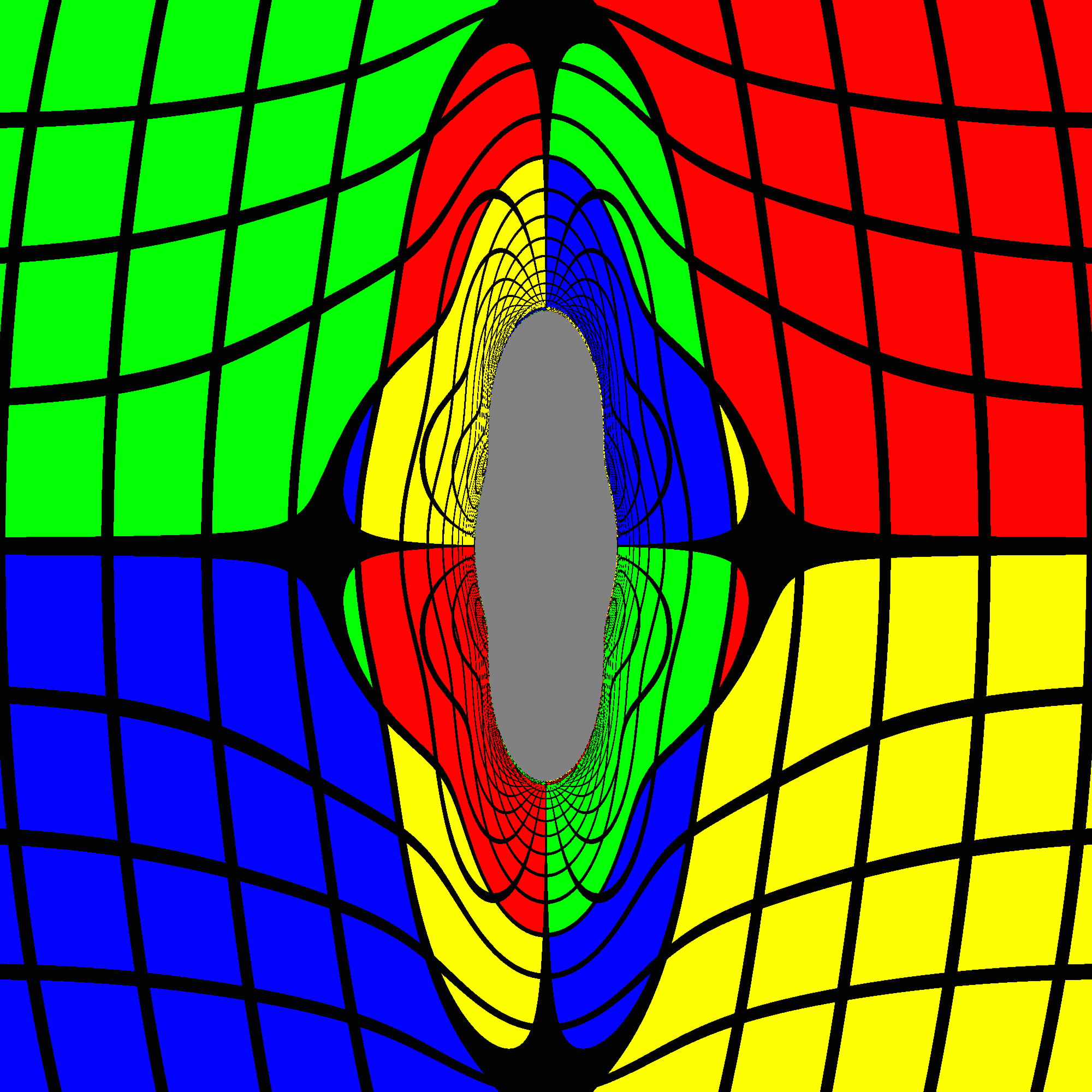}
\par
\includegraphics[width=0.3\textwidth]{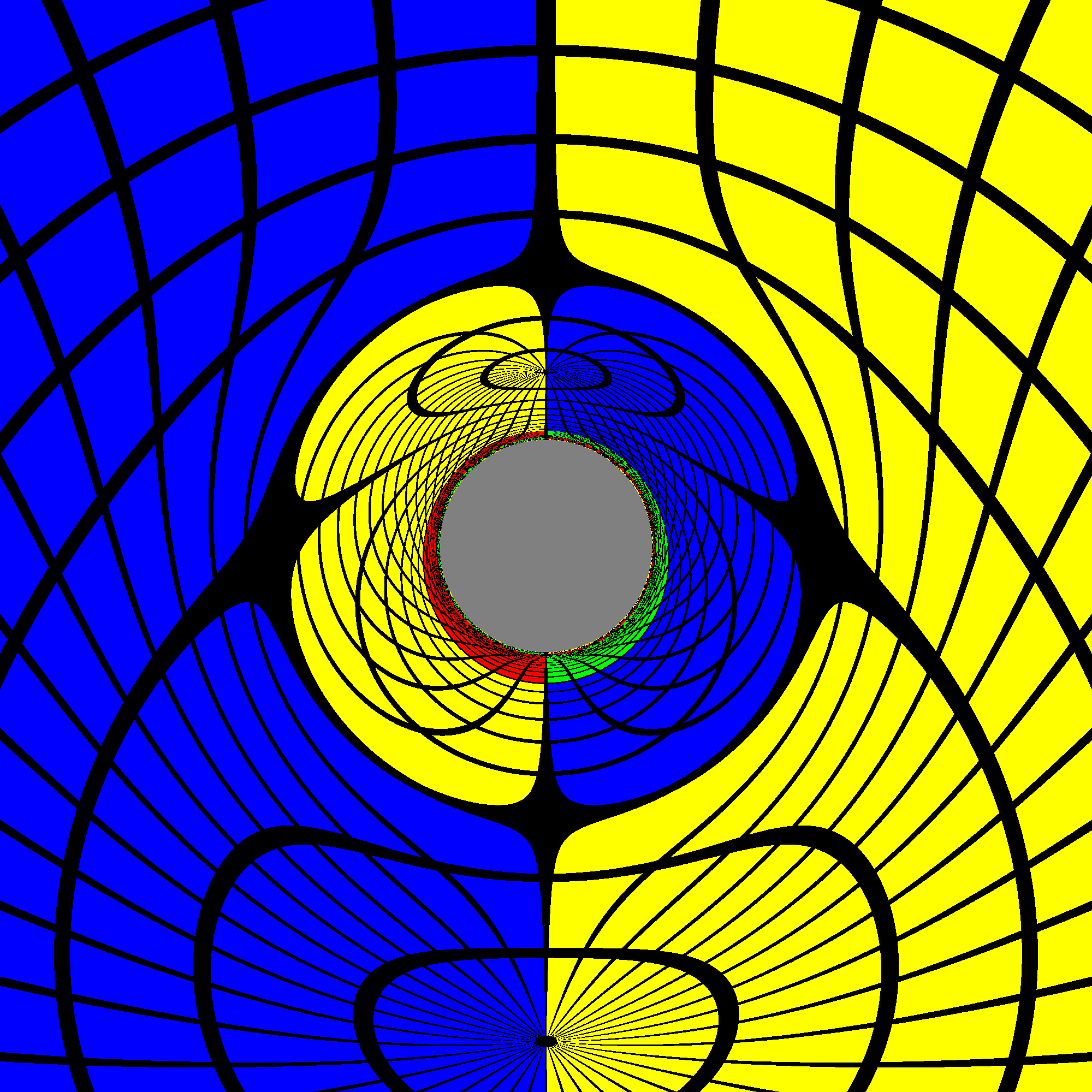}
\includegraphics[width=0.3\textwidth]{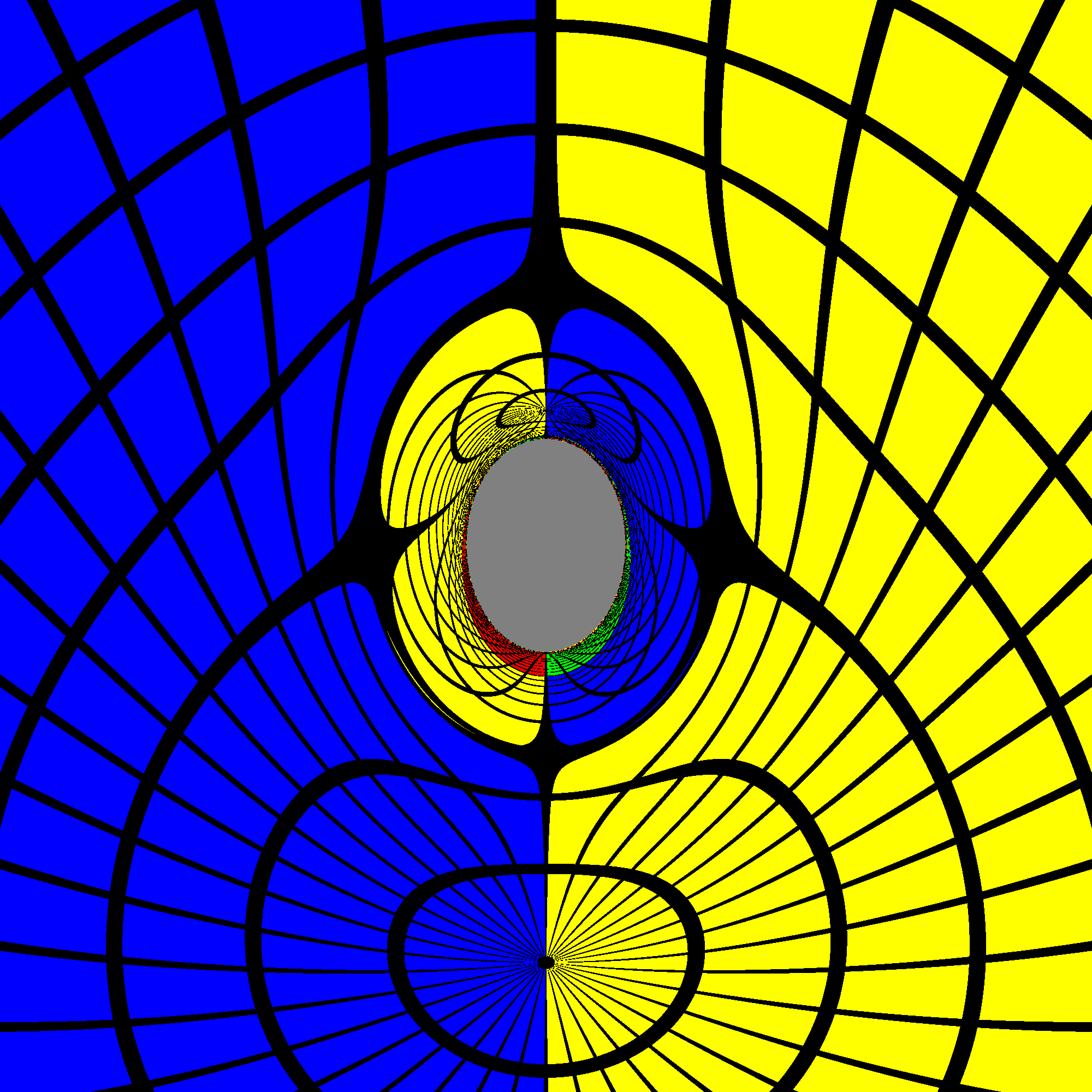}
\includegraphics[width=0.3\textwidth]{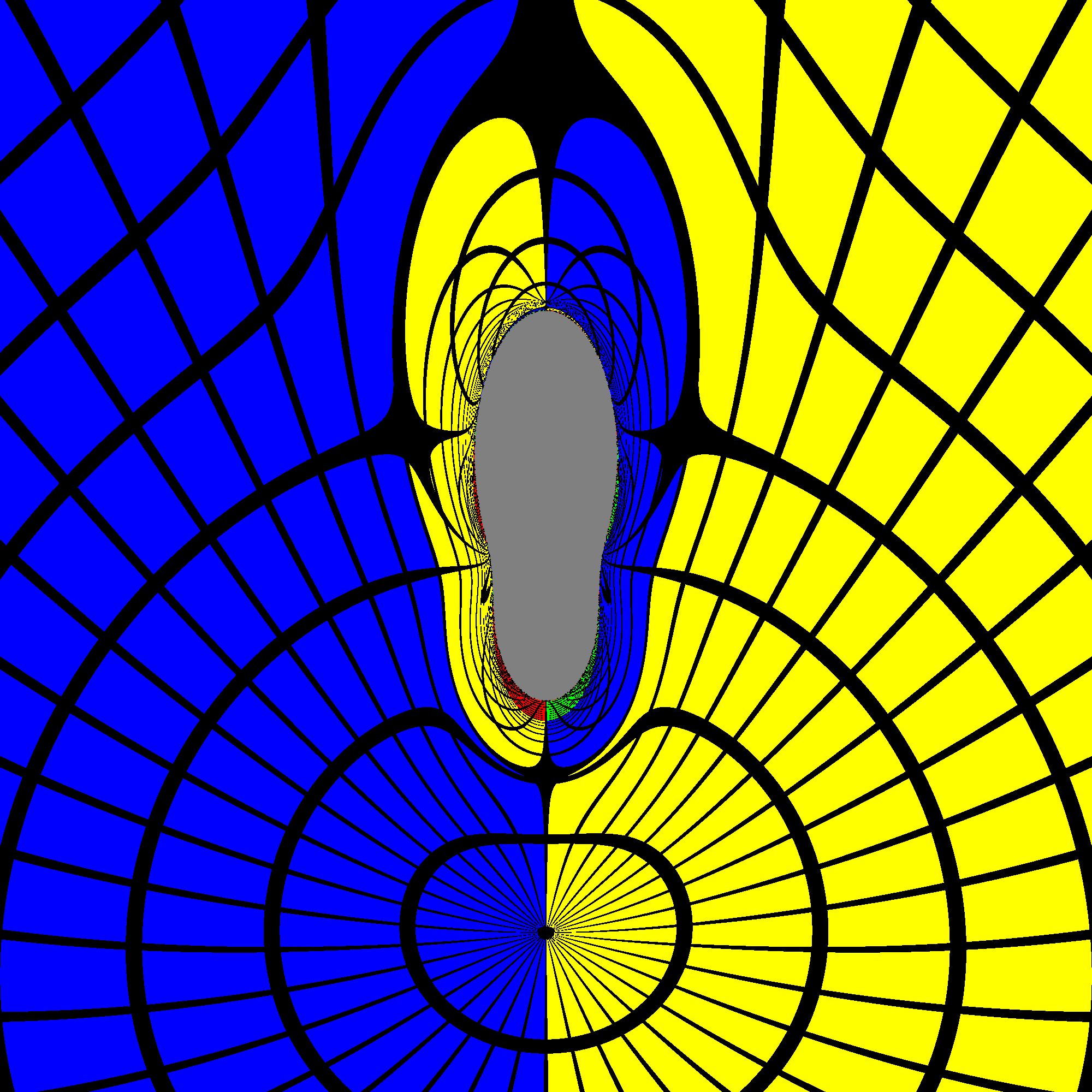} \caption{Celestial sphere
images for a Schwarzschild black hole immersed in BI magnetic fields with
various nonlinearity parameters: $a=0$ (\textbf{Left}), $a=0.5$
(\textbf{Middle}), and $a=2$ (\textbf{Right}). The top and bottom rows
correspond to observer inclination angles $\theta_{o}=90^{\circ}$ and
$\theta_{o}=30^{\circ}$, respectively. The celestial sphere is divided into
four colored quadrants and overlaid with a grid of black lines. Gravitational
lensing distorts the pattern, producing a central black hole shadow (gray
region) surrounded by higher-order images of the celestial sphere. As $a$
increases, the shadow becomes increasingly elongated along the vertical
direction due to nonlinear effects. This elongation diminishes with lower
inclination angles, and the shadow's equatorial symmetry is broken when
$\theta_{o}\neq90^{\circ}$.}%
\label{Fig:=000020CSimages}%
\end{figure}

Fig.~\ref{Fig:=000020CSimages} displays the images of the celestial sphere as
observed for a black hole immersed in BI magnetic fields, highlighting the
effects of various the nonlinearity parameter $a$ and the observer inclination
angle $\theta_{o}$. The columns correspond to $a=0$ (left), $a=0.5$ (middle),
and $a=2$ (right), while the top and bottom rows depict inclination angles of
$\theta_{o}=90^{\circ}$ and $\theta_{o}=30^{\circ}$, respectively. A uniform
color grid on the celestial sphere is distorted by gravitational lensing near
the black hole, forming a central black hole shadow (gray region). A sequence
of higher-order images of the celestial sphere emerges near the shadow edge,
gradually accumulating due to light deflection. As the BI nonlinearity
increases (i.e., with larger $a$), the initially circular black hole shadow
becomes increasingly elongated along the vertical direction. This elongation
is associated with the stretching of closed photon orbits along the $z$-axis
on the meridional plane, as discussed in Fig.~\ref{fig:closed=000020orbits}.
The deformation becomes less pronounced at lower inclination angles, such as
$\theta_{o}=30^{\circ}$, where the observer views the black hole from a more
polar direction. Additionally, while the black hole shadow exhibits reflection
symmetry about the equatorial plane when $\theta_{o}=90^{\circ}$, this
symmetry is broken for $\theta_{o}=30^{\circ}$.

As $a$ increases, the color pattern in the celestial sphere images becomes
more intricate, particularly for $\theta_{o}=90^{\circ}$. For example, in the
first quadrant of the image with $a=0$ and $\theta_{o}=90^{\circ}$, only two
colors---red and blue---are visible. This pattern arises because light rays
are confined to a single plane when $a=0$. In contrast, when $a>0$, four
colors appear in the same quadrant, indicating that some light rays are no
longer restricted to a single plane due to the nonlinear effects of the BI
magnetic field.

\subsection{Accretion Disk}

\label{subsec:Accretion-disk}

We now examine images of a Schwarzschild black hole immersed in BI magnetic
fields and illuminated by a geometrically and optically thin accretion disk
located in the equatorial plane. The observed intensity at specific image
coordinates $\left(  \alpha,\beta\right)  $ is given by
\begin{equation}
I_{o}\left(  \alpha,\beta\right)  =\sum_{m=1}g^{4}\left(  r_{m}\left(
\alpha,\beta\right)  \right)  j_{e}\left(  r_{m}\left(  \alpha,\beta\right)
\right)  ,
\end{equation}
where $r_{m}\left(  \alpha,\beta\right)  $ denotes the radial position of the
$m$-th intersection between the light ray (traced backward from $\left(
\alpha,\beta\right)  $) and the accretion disk. The function $g\left(
r_{m}\right)  $ represents the redshift factor between the observed photon
frequency and the emitted frequency at $r=r_{m}$ \cite{Gralla:2020srx}.
Following \cite{Gralla:2019xty}, light rays with $m=1$, $m=2$, and $m\geq3$
constitute the direct emission, lensing ring, and photon ring, respectively.
For simplicity, we adopt the emission profile
\begin{equation}
j_{e}\left(  r\right)  =\frac{e^{-\left[  \text{arcsinh}\left(  2r\right)
-3/2\right]  ^{2}/2}}{\sqrt{r^{2}+1/4}}.
\end{equation}
The accretion disk is assumed to be rotating: matter outside the Innermost
Stable Circular Orbit (ISCO) follows timelike circular geodesics in the
equatorial plane, while matter inside the ISCO plunges into the black hole,
retaining the same specific energy and angular momentum as at the ISCO. For a
photon with four-momentum $p^{\mu}$, the redshift factor is given by
$g=p_{\mu}u_{o}^{\mu}/p_{\nu}u_{e}^{\nu}$, where $u_{o}^{\mu}$ and $u_{e}%
^{\nu}$ are four-velocities of the observer and the emitter, respectively.
Throughout this paper, the observer is assumed to be static, with $u_{o}^{\mu
}=\left(  1/\sqrt{f\left(  r_{o}\right)  },0,0,0\right)  $. For a
Schwarzschild black hole, the ISCO radius is $r_{\text{ISCO}}=6$. Following
\cite{Gralla:2020srx}, the four-velocity of the emitter outside the ISCO is%
\begin{equation}
u_{e}^{t}=\frac{1}{\sqrt{1-3/r}},\text{ }u_{e}^{\varphi}=\frac{1}{r^{3/2}%
\sqrt{1-3/r}},
\end{equation}
while inside the ISCO, it is given by
\begin{equation}
u_{e}^{t}=\frac{2\sqrt{2}}{3}\frac{1}{1-2/r},\;u_{e}^{r}=-\frac{1}{3}\left(
\frac{6}{r}-1\right)  ^{3/2},\;u_{e}^{\varphi}=\frac{6}{\sqrt{3}r^{2}}.
\end{equation}

\begin{figure}[ptb]
\begin{centering}
\includegraphics[width=0.3\textwidth]{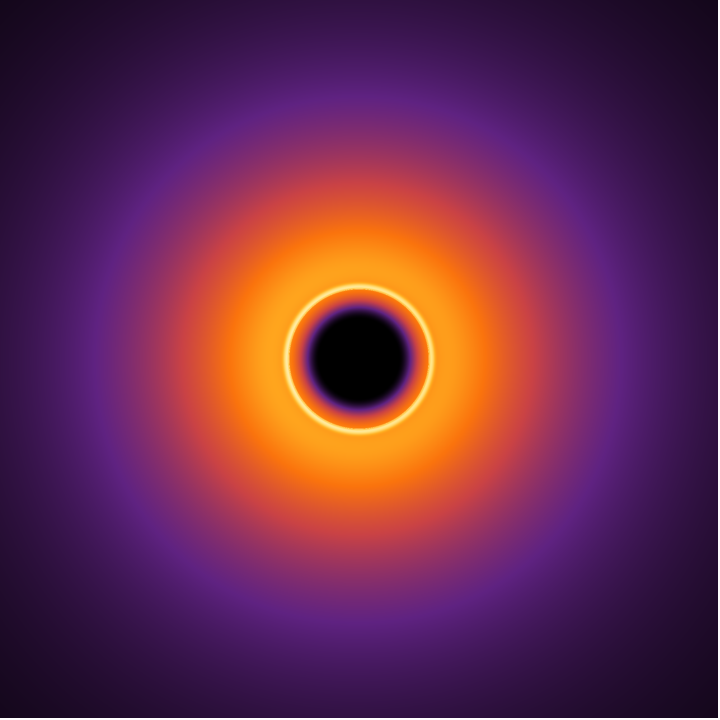} \includegraphics[width=0.3\textwidth]{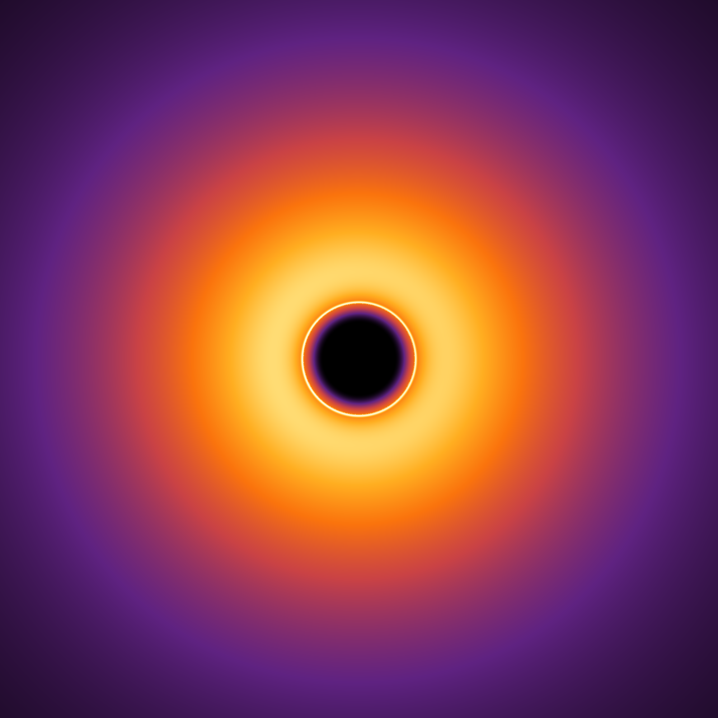}
\includegraphics[width=0.3\textwidth]{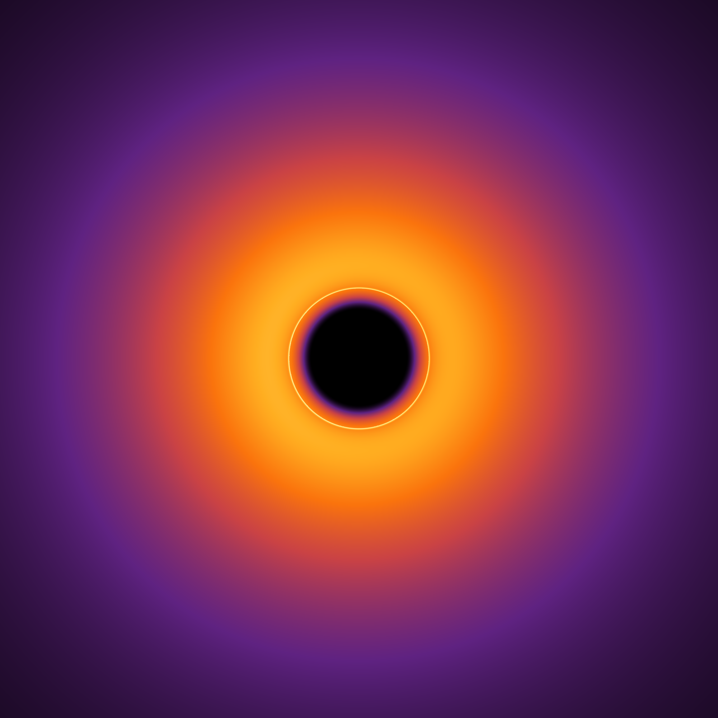}
\par\end{centering}
\begin{centering}
\includegraphics[width=0.3\textwidth]{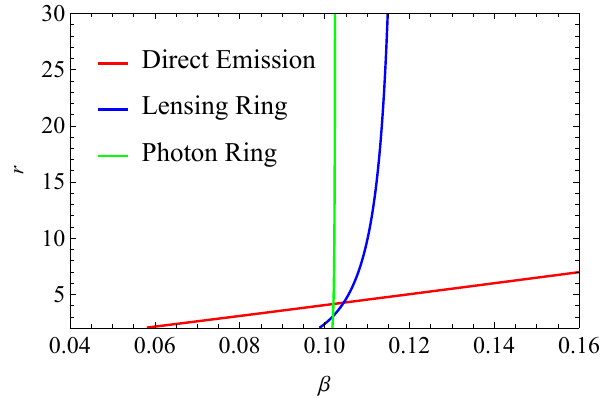} \includegraphics[width=0.3\textwidth]{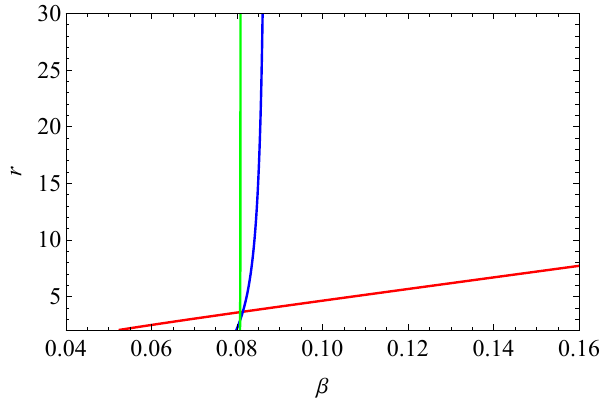}
\includegraphics[width=0.3\textwidth]{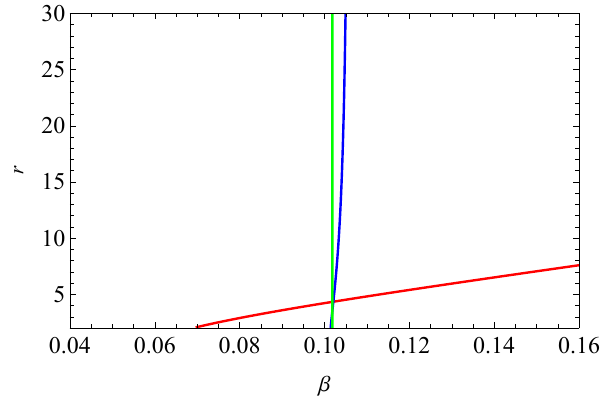}
\par\end{centering}
\caption{\textbf{Top Row}: Images of a rotating accretion disk around a
Schwarzschild black hole immersed in BI magnetic fields, viewed by a face-on
observer at $\theta_{o}=0^{\circ}$, for nonlinearity parameters $a=0$
(\textbf{Left}), $0.5$ (\textbf{Middle}), and $2$ (\textbf{Right}). Each image
displays a circularly symmetric disk-like emission region, a central shadow,
and a bright lensing ring. The sizes of the shadow and lensing ring vary
non-monotonically with $a$. \textbf{Bottom Row}: The corresponding transfer
function $r_{m}\left(  0,\beta\right)  $ for the direct emission $\left(
m=1\right)  $, lensing ring $\left(  m=2\right)  $, and photon ring $\left(
m\geq3\right)  $. The lensing ring becomes progressively narrower with
increasing $a$.}%
\label{Fig:=000020Diskimagedeg0}%
\end{figure}

The top row of Fig. \ref{Fig:=000020Diskimagedeg0} shows the images of the
rotating accretion disk as viewed by a face-on observer at $\theta
_{o}=0^{\circ}$, for nonlinearity parameters $a=0$, $0.5$, and $2$. For this
viewing angle, the images exhibit circular symmetry. Each image reveals a
disk-like direct emission region surrounding a central shadow, overlaid by a
thin, intensely bright lensing ring. Notably, the sizes of both the shadow and
the lensing ring vary non-monotonically with $a$: they are largest for $a=2$,
intermediate for $a=0$, and smallest for $a=0.5$. This non-monotonic behavior
is closely associated with the compression and stretching of bound photon
orbits along the $x$-axis on the meridional plane for $a=0.5$ and $a=2$,
respectively. The bottom row of Fig.~\ref{Fig:=000020Diskimagedeg0} plots
$r_{m}\left(  0,\beta\right)  $ as a function of $\beta$, referred to as the
transfer function, for the direct emission $\left(  m=1\right)  $, lensing
ring $\left(  m=2\right)  $, and photon ring $\left(  m\geq3\right)  $. Both
rows demonstrate that the lensing ring becomes significantly narrower as $a$ increases.

\begin{figure}[ptb]
\includegraphics[width=0.3\textwidth]{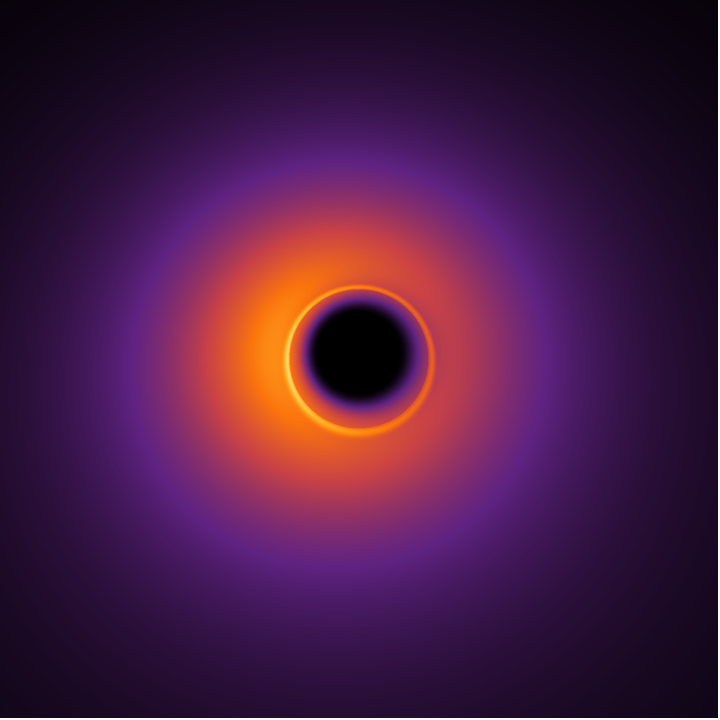}
\includegraphics[width=0.3\textwidth]{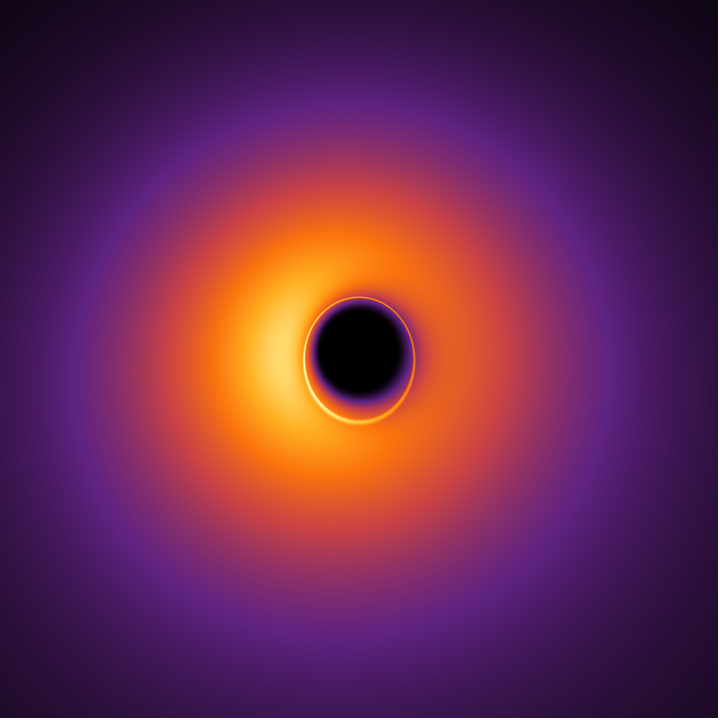}
\includegraphics[width=0.3\textwidth]{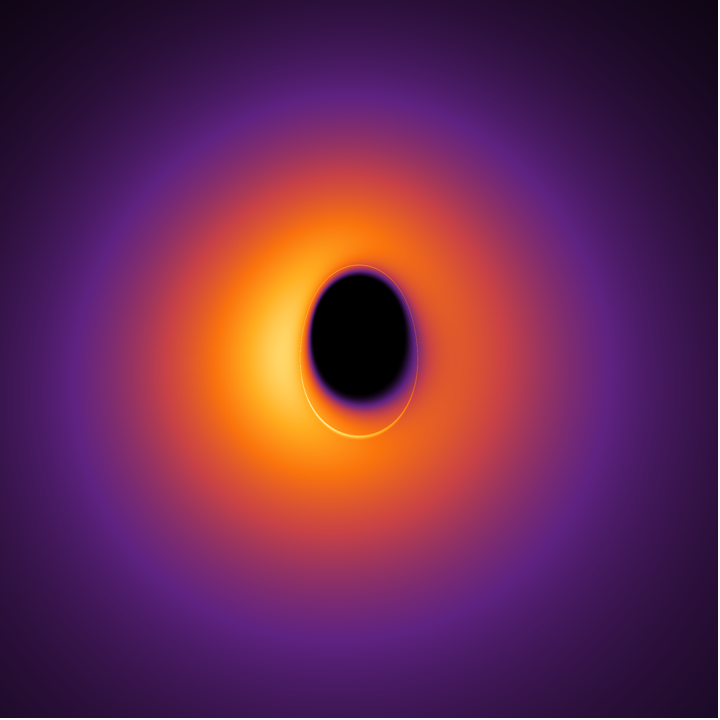}
\par
\includegraphics[width=0.3\textwidth]{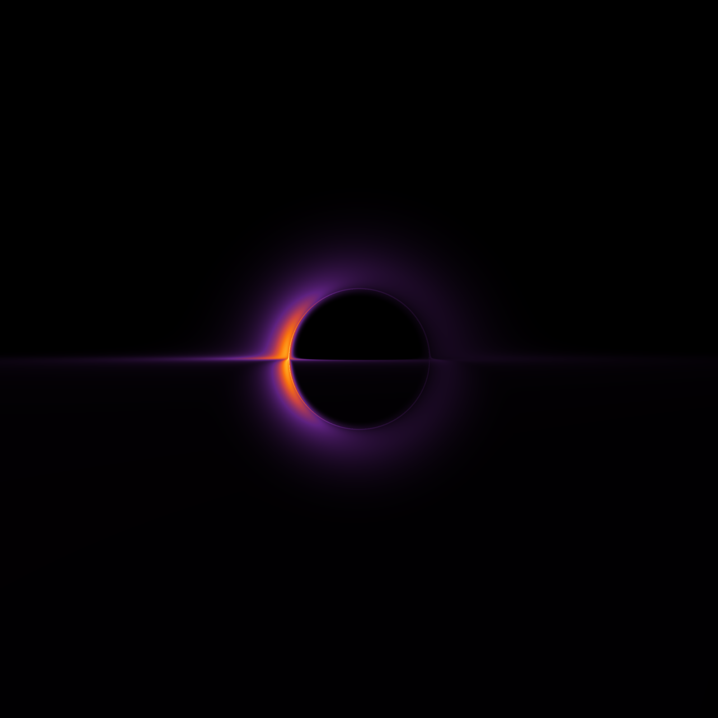}
\includegraphics[width=0.3\textwidth]{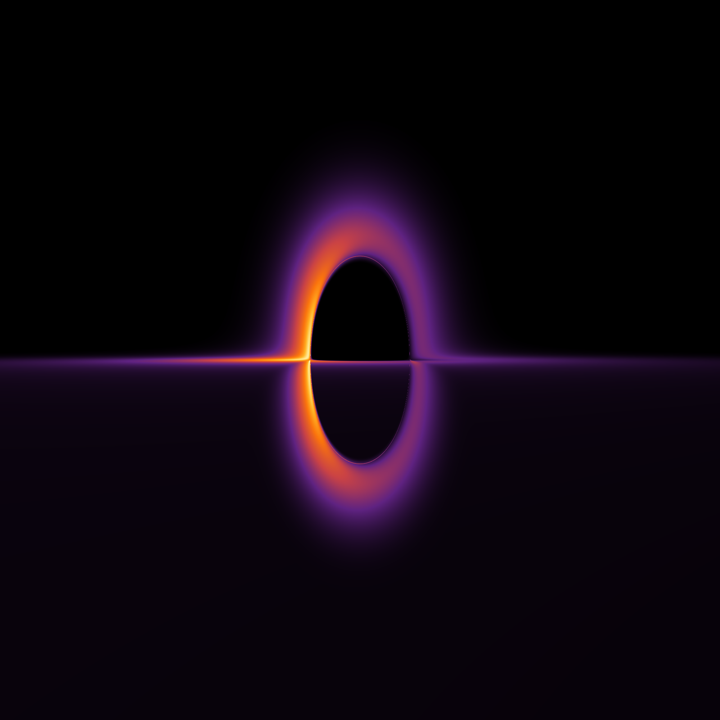}
\includegraphics[width=0.3\textwidth]{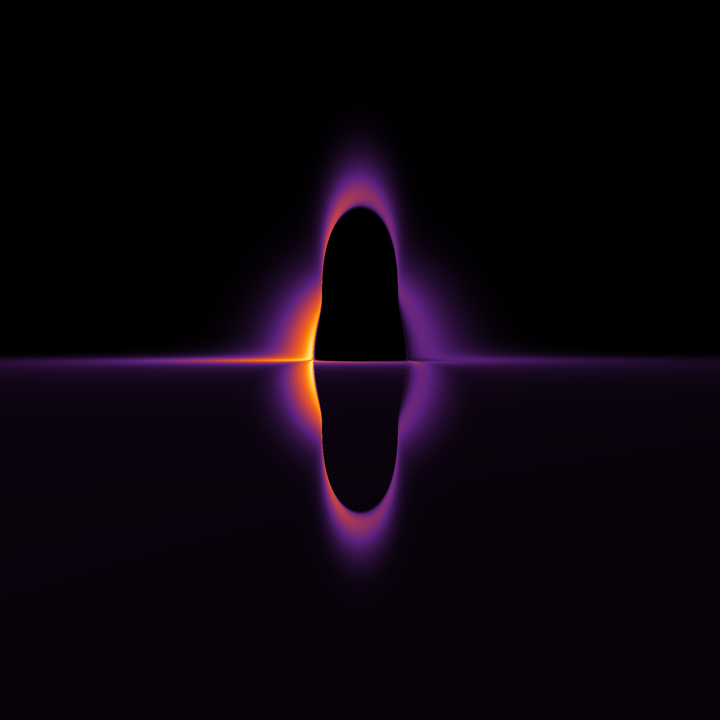}\caption{Images of
a rotating accretion disk viewed at inclination angles $\theta_{o}=17^{\circ}$
(\textbf{Top}) and $\theta_{o}=89^{\circ}$ (\textbf{Bottom}), for nonlinearity
parameters $a=0$ (\textbf{Left}), $0.5$ (\textbf{Middle}), and $2$
(\textbf{Right}). For $\theta_{o}=17^{\circ}$, the black hole shadow remains
nearly circular, though a gradually increasing vertical elongation emerges
with increasing $a$. The lensing ring remains narrow and bright, and becomes
thinner for larger $a$. For $\theta_{o}=89^{\circ}$, the images appear
significantly dimmer due to reduced direct emission, which is confined to a
thin horizontal stripe and the upper portion of the bright ring. The lensing
ring forms the lower portion of the bright ring. The vertical elongation of
the shadow becomes more pronounced at this high inclination.}%
\label{Fig:=000020Diskimagesdeg17and90}%
\end{figure}

Fig. \ref{Fig:=000020Diskimagesdeg17and90} displays the images of the rotating
thin accretion disk as viewed at inclination angles $\theta_{o}=17^{\circ}$
(top row) and $\theta_{o}=89^{\circ}$ (bottom row), respectively. The disk's
rotation induces a pronounced Doppler effect, resulting in a clear left-right
asymmetry: emission from matter approaching the observer (left side) is
blueshifted and significantly brighter, while emission from receding matter
(right side) is redshifted and dimmer. At $\theta_{o}=17^{\circ}$, the shadow
boundary shows a slight but increasingly noticeable deviation from circularity
as $a$ increases. This deformation manifests as a vertical elongation of the
shadow aligned with the magnetic field direction, consistent with the
stretching of closed orbits in the meridional plane. The lensing ring
maintains a narrow, high-intensity structure that becomes even thinner with
increasing $a$. At the near-edge-on inclination of $\theta_{o}=89^{\circ}$,
the total observed flux is significantly reduced. The direct emission consists
of a thin horizontal bright stripe across the center and the upper portion of
the bright ring surrounding the shadow. In contrast, the lensing ring
primarily contributes to the lower portion of the bright ring. As $a$
increases, the vertical elongation of the shadow becomes more pronounced
compared to the lower inclination case.

\section{Conclusion}

\label{sec:Conclusion}

In this paper, we first investigated the configuration of an asymptotically
uniform BI magnetic field in the fixed background of a Schwarzschild black
hole. Our analysis shows that nonlinear electromagnetic effects become
significant near the black hole horizon, particularly in the polar regions. As
the nonlinearity parameter $a$ increases, the magnetic field exhibits a
noticeable enhancement near the poles, accompanied by a higher density of
field lines in this region. This contrasts with the Maxwell case, where the
field remains relatively uniform near the black hole's polar axis.

We also analyzed the motion of light rays in the effective geometry induced by
BI magnetic fields. While closed photon orbits remain circular on the
equatorial plane, they become prolate on the meridional plane, displaying
elongation along the polar axis. This stretching correlates with the
amplification of the magnetic field magnitude in the polar regions.
Furthermore, we simulated black hole images and shadows in the presence of BI
magnetic fields. Our simulations show that the shadow, which is circular in
the Maxwell limit, becomes increasingly elongated along the polar direction as
$a$ increases, particularly for large observer inclination angles.

This paper highlights the importance of considering nonlinear electrodynamics
in strong gravitational backgrounds, where observable signatures can deviate
substantially from predictions based on the Maxwell theory. The distinctive
deformations of black hole shadows presented here may serve as potential
probes for future high-resolution astronomical observations
\cite{Johnson:2019ljv,Himwich:2020msm,Gralla:2020srx}. Future research will
explore these phenomena in the context of rotating black holes, investigate a
wider range of the nonlinearity parameters and alternative emission models,
and aim to compare theoretical predictions with observational data from
facilities like the EHT.

\begin{acknowledgments}
We are grateful to Lang Cheng and Tianshu Wu for useful discussions and
valuable comments. This work is supported in part by NSFC (Grant Nos.
12347133, 12250410250, 12275183 and 12275184) and Discipline Talent Promotion
Program of /Xinglin Scholars (Grant No. QNXZ2018050).
\end{acknowledgments}

\bibliographystyle{unsrturl}
\bibliography{ref}

\end{document}